\documentclass{article}[12pt]
\usepackage{amsmath}
\usepackage{amssymb}
\usepackage{amsthm}
\usepackage{amsfonts}
\usepackage[dvips]{graphicx}

\def\abstract#1{\vskip 7mm
        \begin{center}{\large Abstract}\par \smallskip
                \begin{minipage}[c]{12cm}
                        \small #1
                \end{minipage}
        \end{center}
}

\def\title#1{\begin{center}{\Large\bf #1}\end{center}}
\def\author#1{\vskip 5mm \begin{center}{#1}\end{center}}
\def\address#1{\begin{center}{\it #1}\end{center}}

\newcommand{\ssmatrix}[4]%
{\begin{pmatrix} #1 & #2 \\ #3 & #4 \end{pmatrix}}
\makeatletter
\@ifundefined{lesssim}{}{}
\@ifundefined{gtrsim}{}{}
\def\vereq#1#2{\lower3pt\vbox{\baselineskip1.5pt \lineskip1.5pt
\ialign{$\m@th#1\hfill##\hfil$\crcr#2\crcr\sim\crcr}}}
\makeatother

    {\mbox{}\hfill\mbox{\rm\normalsize$\Box$}\\%
    }

\newcommand{\EPS}{{\cal E}}

\newcommand{\nn}{\nonumber \\ }

\begin{document}

\title{%
Head-on collision and merging entropy of black holes:
reconsideration of Hawking's inequality}
\author{%
  Masaru Siino\footnote{E-mail:msiino@th.phys.titech.ac.jp} 
}
\address{%
  Department of Physics, Tokyo Institute of Technology, \\
  Oh-Okayama 1-12-1, Meguro-ku, Tokyo 152-8550, Japan  
}

\abstract{
We evaluate how much energy can be converted into gravitational radiation in head-on collision of black holes.
We estimate it by the area theorem of black hole horizon incorporating merging entropy of colliding black holes from a viewpoint of black hole thermodynamics. Then we obtain an upper bound of energy ratio of the gravitational radiation which is smaller than the upper bound originally derived by Hawking.
The fact that this estimation is not inconsistent with the results of both numerical investigations in low- and high-energy head-on collision implies that thermodynamics of coalescing black holes requires the contribution of the merging entropy.}


\section{Introduction}
In head-on collision of black holes, evaluation of energy converted into gravitational radiation is an important aim in astrophysical application of general relativity.
The leading work is made by Hawking\cite{SH} using an area theorem of black hole event horizon, which placed an upper limit of 29\% on the total energy radiated when two black holes initially at rest coalesce, in head-on collision of spin-less black holes with identical masses. Using similar argument based on the area theorem, Penrose\cite{RP} derived an upper bound of 29\% for ultra-relativistic head-on collision.
However numerical simulation later showed that the gravitational radiation is far less than expected. They reported true value of energy ratio of the gravitational radiation around 0.1\% \cite{NA}. Furthermore the numerical result can be accounted for the perturbative analysis\cite{PRP}, while one may get the impression that such calculations based on the area theorem are not suitable.
Indeed, the fact that the gravitational wave barely radiates from the system, is the evidence of Hawking's upper bound, but if the upper bound were a result of sufficiently detailed evaluation, it would be expected to be closer to the numerical result and to provide any prediction. In this sense, the gravitational radiation is far from enough to feel such a mathematical analysis is useful.
Why is there not much gravitational radiation in this simple head-on collision?

The essence of Hawking's discussion is the area theorem of black hole event horizon\cite{HS}. The area theorem is proved, based on the mathematical concepts about event horizon generators, which are the Raychaudhri equation and their future completeness.
Since these mathematical concepts are fundamental, Hawking's bound may not be severe but will be universal independently of detailed physical situations.
Indeed, for the numerical experiments\cite{NA}, Hawking's bound suffers no problem but says nothing instructive.
Then if we want to get any useful information from such a mathematical discussion, we should change the bound more severe by making the discussion more precise.
In the present work, we will make the upper bound smaller by revising the application of the area theorem to the black hole collision.

The work of Bekenstein\cite{BS} and Hawking\cite{HS} has shown, and that of many others confirmed, that the mechanical laws governing classical systems containing black holes can be placed in analogy with those of thermodynamics.
The area theorem is one of the important theorems of the black hole mechanics.
Its importance have increased since it is included in the black hole thermodynamics as increasing low of the Bekenstein-Hawking entropy by the assumption that the stationary black hole possesses entropy which is a fourth of its horizon area.
If even in dynamical situation the black hole entropy should be generally given by a fourth of the horizon area, thermodynamical estimation of the entropy increasing will provide a dynamical changing of the horizon area, which will be evaluated for calculating the energy ratio of gravitational radiation.
Then in the present study of black hole collision, we expect any contribution of additional entropy like the entropy increase of mixing for the merging process, which was not incorporated in the Hawking's upper bound\cite{SH}.
The entropy increase $\Delta S$ should be added to a black hole area inequality as extra area increase $\Delta A$ of black hole by $\Delta A=4\Delta S$ and would gives smaller upper bound of the energy ratio. 

Recently in high-energy head-on collision of spin-less black holes with identical masses, in which the colliding black holes initially possess large relative velocity, a remarkable numerical
investigation\cite{HC} has shown that the large amount of energy is converted into gravitational radiation.
Its result suggests that in the maximally high-energy limit, the energy ratio tends to 14\% of total energy in extrapolation by the dependence of $\gamma$ fitting the prediction\cite{HC} of linear perturbation based on the zero frequency limit and point particle approximation, that is a half of Hawking's and Penrose's bounds, and quite close to the estimation 16\% of D'Eath, Payne\cite{DP} using perturbative technique.
Now, our problem has become more complicated than before.
Can we explain both why the energy ratio of the gravitational radiation is so small in the low-energy collision and is so large in the high-energy collision?

The purpose of the present article is to find an estimation of the energy ratio of the gravitational radiation in the head-on collision of black holes, based on the area theorem of black hole horizon incorporating merging entropy. In order to insist the contribution of the merging entropy, we aim for an evaluation of the energy ratio which is in a good agreement with the numerical\cite{NA}\cite{HC} and perturbative \cite{PRP}\cite{DP} results. 
In the second section, we recall the original work of Hawking. On the top of the third section, we give a key idea of the present study.
Our discussion puts a base on the issue of black hole thermodynamics.
Furthermore in the third section, thermodynamical investigation for merging entropy is shown. Using the merging entropy the forth section provides an evaluation of the energy ratio of the gravitational radiation in the low- and high-energy head-on collision of spin-less black holes with identical masses. 
The final section is devoted to conclusion and discussions.
We choose units so that $G=c=\hslash=1$.

\section{Hawking's bound}
When two black holes coalesce, the amount of energy converted into the gravitational radiation is restricted by Hawking's simple discussion based on the area theorem of black hole event horizon. Here we recall the original discussion in Reference\cite{SH}.

We simply consider two spin-less black holes with mass $M_1$ and $M_2$.
In head-on collision of the spin-less black holes, the final state will be also a spin-less black hole with mass $M_{tot}$. In Hawking's discussion, it is supposed that colliding black holes are initially far separated. Then each black hole is considered to be an isolated Schwarzschild black hole with its own ADM-mass $M_i$.
The energy $E$ converted into the gravitational radiation is related to the mass parameters by the energy conservation 
\begin{align}
M_1+M_2=M_{tot}+E.
\label{eqn:EC}
\end{align} 
On the other hand, we have the area theorem of black hole event horizon\cite{AT} which states that the area of the event horizon never decreases between two not intersecting spatial hypersurfaces under certain conditions. Then we have an inequality between initial and final area of black hole horizon,
\begin{align}
M_1^2+M_2^2\leq M_{tot}^2,
\label{eqn:AT}
\end{align}
if we suppose that initial horizon area is the sum of areas of two isolated black holes whose horizon areas are $16\pi M_i^2$.
The equality would be attained if two stationary black holes instantaneously merged into a final stationary black hole.
Consequently, the relations (\ref{eqn:EC}) and (\ref{eqn:AT}) restrict the energy converted into the gravitational radiation as
\begin{align}
E<M_1+M_2-\sqrt{M_1^2+M_2^2}.
\end{align}
Especially in the case of identical masses $M_1=M_2$, the energy ratio of the gravitational radiation $E/2M_1$ should be less than $29\%$.

After that work, numerical simulations\cite{NA} have shown in the identical mass head-on collision the energy ratio of the gravitational radiation ($\sim$0.1\%) is two orders of magnitude smaller than Hawking's bound.
Anyway, that means the evaluation of the area theorem by Schwarzschild black hole horizon is not incorrect.
On the other hand, the upper bound is too large to predict anything for the collision. In a sense, it is not realistic to evaluate the event horizon area of colliding black hole by its gravitational mass as $16\pi M_i^2$, since the event horizon is determined by not only local geometry around the black hole but also global geometry including the other black hole. 
How can we turn the upper bound severe by improving the evaluation of the event horizon area?
It is the main aim in the present article to clarify why gravitational radiation is not much in such head-on collision of black holes, by our improved evaluation of the area theorem in inequality (\ref{eqn:AT}).


\section{Our estimation}

The key idea of the present work is that the event horizon area of just merging black hole is far different from that of Schwarzschild horizon.
Now we must explain that the horizon area of dynamical black holes is larger than the sum of the areas of two isolated black holes.
Roughly speaking, the event horizon (EH) is extended from the Schwarzschild horizon to the direction of the collision, since a light ray emanating into this direction will be affected also by the gravitation of the opponent black hole. 
It becomes harder for the light ray to escape from the gravitation to the future null infinity.

\begin{figure}[hbtp]
\includegraphics[width=9cm]{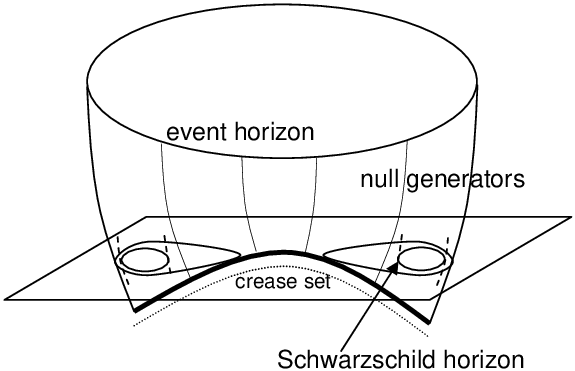}
\caption{ An event horizon of the coalescing black holes is illustrated. The event horizon is extended into the direction of the collision. Moreover a line-like crease set appears.}
\label{fig:sp}
\end{figure}

Is this effect significant even if the two black holes are well separated?
Here we are noticed that the situation depends on the timeslicing we choose. 
The event horizon of the colliding black holes is illustrated in Fig. \ref{fig:sp}. 
From this figure, we see the deformation of the horizon is remarkable near spacetime point where two black holes merge.
If we choose an initial hypersurface which includes the region near the merging spacetime point, the area of the event horizon is much larger than the sum of the area of two Schwarzschild black holes even on the initial hypersurface.
Nevertheless, since we choose the initial hypersurface which does not contain that region, each black hole seems to be a Schwarzschild one and the area of the horizon is approximately the sum of the Schwarzschild horizon area.

It might sound as if the initial states are selected such that the part of them is always in highly dynamical regime. 
Even if almost stationary initial states are selected, however, that is not the case. If the black holes merge far in the future then the merging spacetime point will be in a region spatial to the initial black holes, since the event horizon is achronal boundary\cite{AT}\cite{MS1}.
Therefore we could change the initial hypersurface to that contains the merging point even with almost stationary initial state\footnote{Similar situation is discussed for the binary black holes in \cite{SIda}}.
From this viewpoint, we realize that the horizon area increases during the dynamical process before the two black holes merge.

In other words, there are two contributions to the correction of the area from the Hawking's discussion for the deformation of the event horizon.
One is the correction for the deformation which is already exists even when two black holes are far separated on the initial hypersurface.
Nevertheless, this contribution would be strongly suppressed by their separation since such a gravitational interaction of the two black holes might be in order of inverse-square of their separation when two black holes are well separated.
Another is the correction which occurs when two black holes getting near.
Since this would be mainly caused by the black hole occultation which is proportional to the cross-section over $4\pi r^2$, will become significant suddenly when two black holes are getting close.
Since the area of event horizon increases, the lower bound of area after coalescence should be larger than the Hawking's estimation $16\pi(m_1^2+m_2^2)$ by the deformation of the area $\Delta A_i>0$.



The difference $\Delta A_i$ between EH area and $16\pi M_i^2$, which is the area of a corresponding Schwarzschild black hole with mass $M_i$, should be included into the area theorem eq.(\ref{eqn:AT}) as
\begin{align}
M_1^2+M_2^2+\frac{\Delta A_1+\Delta A_2}{16\pi}\leq M_{tot}^2.
\label{eqn:AT2}
\end{align}

For simplicity, we consider head-on collision of spin-less black holes with identical masses $M_1=M_2$, which finally settles to a Schwarzschild black hole with mass $M_{tot}$.
Since two black holes are identical, the excesses of the EH area over the Schwarzschild horizon area are also identical $\Delta A_{tot}=\Delta A_1+\Delta A_2=2\Delta A_1$.
From the conservation low $M_1 + M_2=M_{tot}+E$ and the area theorem eq.(\ref{eqn:AT2}), we obtain another upper bound of the energy ratio,
\begin{align}
 2M_1^2+\frac{1}{8\pi}\Delta A_1&\leq  (2M_1-E )^2\\
\Rightarrow\ \ \ \frac{E}{2M_1}&\leq 1-\sqrt{\frac12 +\frac{\Delta A_1}{32\pi M_1^2}},
\label{eqn:AT3}
\end{align}
which is lowered if EH area is larger than $16\pi M_i^2$ ($\Delta A_i>0$). Of course, that is expected under certain physical conditions.

Nevertheless, as depicted in Figure \ref{fig:sp} this area difference is highly dynamical and non-linear, so that analytical decision will be very difficult.
Then in the present article, we will estimate the area difference between EH in the black hole coalescence and Schwarzschild horizon, from black hole thermodynamical discussion.
To consider thermodynamical aspects of the black hole coalescence, we adopt an analogy with a gas cylinder in uniform gravitation as a toy model of a merging black hole in the following subsection.

\subsection{merging entropy}

The work of Bekenstein\cite{BS} and Hawking\cite{HS} has shown, and that of many others confirmed, that the mechanical laws governing classical systems containing black holes can be placed in analogy with those of thermodynamics.
Further, the resulting correspondence between mechanical black hole variables (horizon area, surface gravity, etc.) and thermodynamic variables (entropy, temperature, etc.) has independent physical meaning when quantum mechanics is taken into account. This correspondence has been made explicit for a number of examples.
Especially the important relation for the present work is that of the entropy and horizon area, $S=A/4$.

An upper bound of gravitational radiation for coalescence of black holes is derived by Hawking\cite{SH} based on the area theorem of the black hole horizon\cite{AT}.
On the other hand, from the viewpoint of black hole thermodynamics the area theorem plays the role of the second law of thermodynamics, that is, the area theorem insists the entropy increases in the transition from an initial stationary black hole to a final stationary 
black hole\footnote{To justify the black hole thermodynamics, we must invoke the existence of 
quantum field. Nevertheless, for such a massive black hole considered in the numerical
simulations its contribution to the evolution equation is negligible, and we can 
consider the first low of the black hole thermodynamics only as a relation between the
horizon area and black hole entropy.}.

On the stationary spin-less black hole with mass $M$, its EH coincides with Schwarzschild horizon whose area will be $16\pi M^2$ in Schwarzschild coordinate.
Nevertheless when two black hole EHs are just merging, the black holes are no longer than stationary and their EHs will be different from the Schwarzschild horizon, so that the EH area of a black hole with mass $M_1$ is estimated by $16\pi M_1^2+\Delta A_1$ in which $\Delta A_1$ will be brought by the gravitation of the other black hole with mass $M_2$.

From the viewpoint of black hole thermodynamics, the area difference $\Delta A_i$ corresponds to entropy increase $\Delta S_i=\Delta A_i/4$ between an initial stationary black hole and a final stationary black hole.
In Hawking's discussion, the total area was estimated by the summation of each area of individual Schwarzschild black hole as $16\pi(M_1^2+M_2^2)$. 
That may thermodynamically imply to omit the increasing of the entropy during the mixing process.
In other words, this is to assume that two stationary merging black holes instantaneously settle into one final stationary black hole without changing its intrinsic configuration. The implication of the black hole uniqueness, however, that the final stationary black hole has forgotten its formation history means that the final black hole has suffered any mixture of the intrinsic configuration, and then there would be any entropy increase in this merging process. 
If we stands on the viewpoint that the total entropy is divided into two black holes, the mixing contribution is also divided and causes the corresponding corrections of the areas of the event horizons.
Therefore we expect that we are able to evaluate area difference $\Delta A_1$ from the entropy increase $\Delta S_1$.

Though the above discussion resembles to that of Gibbs' paradox for the mixing entropy, we are careful about difference between the ordinary mixing entropy and the present increase of black hole entropy.
In Gibbs' paradox if two substances are identical, merged system immediately becomes equilibrium and there is no entropy increase of mixing.
On the other hand for the black hole coalescence, even if the two black holes are identical, merged black hole does not settles to stationary (equilibrium) state until it forgets its own formation history, and then will suffer an entropy increase which we call merging entropy of the black holes.

The way to calculate such an entropy increase should be essentially based on state counting in statistical mechanics. That microscopic nature of a black hole, however, has not been fully clarified yet. Especially, since this system is highly dynamical, it is unlikely that we can evaluate the entropy of colliding black holes, by the microscopic calculation.
Rather an easy way of estimating the entropy increase upon the merging (mixing) of black holes may be to consider a toy model of a merging (mixing) thermodynamical black hole affected by the gravitational force of the other black hole.
Here we had better emphasize that the choice of model would not affect the essential factors except for the numerical coefficients.
Indeed, in the following, we see that the thermodynamical relation between merging entropy and gravitational acceleration is similar to the relation handled in the discussion of entropic force\cite{V}.
Since in the present article we aim to fit our estimation to the numerical and perturbative results, we choose a convenient model without verifying its inevitability.

By way of example, the entropy of mixing for gas in chambers may be calculated by Gibbs' Theorem which states that when two different substances mix, the entropy increase upon mixing is equal to the entropy increase that would occur if the two substances were to expand alone isothermally into the mixing volume\footnote{In this sense, the term "entropy of mixing" is a misnomer, since the entropy increase is not due to any "mixing" effect.}.
As a working hypothesis, we suppose that the merging entropy of black holes is also evaluated by similar individual isothermal expansion of any substances into a mixing volume.

The entropy increase of merging black holes would be brought by a black hole getting near the opponent black hole, since we speculate that its substance between two black holes diffuses into the direction of the opponent black hole losing its binding force by cancellation of gravitational acceleration.
Here we intuitively assume that the substances are bound not by wall rather by gravitational force around the black hole. 
Then we could estimate the merging entropy for black holes by calculating the entropy difference of diffusing matter between the state with gravitation and without gravitation.
For simplicity we will assume that the entropy increase occurs isothermally in an appropriate temperature, though it might not be valid to apply isothermal process to such a merging black hole.\footnote{For the present, we could not think of any further justification for the 
isothermal assumption in the mixing of gases.
Nevertheless, 
since the entropy increase is variable determined by initial and final states not by intermediate process, any appropriate combination
of isothermal process and adiabatically cooling process may realize this
dyanamical process. If so, one would be possible to fix the parameter $alpha$
further.}

As a simplest toy-model, we imagine that the substances composed of ideal gas in the gravitational potential related to the black holes.
Instead of expanding into the mixing volume, a substance will diffuse into intermediate space between the two black holes by balance of gravitation of two colliding black holes. Since that is one-dimensional force balance, in a toy model we simply imagine uniform gravitational force caused by each black hole. 
Around one of colliding black holes, initially there is no other black hole and at coalescence the other black hole appears, so that two uniform gravitational forces caused by the two black holes with equal strength and opposite directions, offset each other. 
Losing their binding force, their substances diffuse into the intermediate space between the two black holes.

Suppose a gas cylinder with vertical length $L$ (to be a distance between the two black holes) and with area of section $s$, is in a uniform gravitation.
We will estimate the entropy increase by calculating the entropy difference between the ideal gases in the uniform gravitation and not in. 

The chemical potential at height $h$ of an ideal gas in a uniform gravitational acceleration $g$ and temperature $\tau=k_B T$ is given by
\begin{align}
\mu(h,\tau)=\tau\log\frac{n(h,\tau)}{n_q}+mgh,
\end{align}
where $m$ is the mass of  particle and $n(h,\tau)$ is its number density at height $h$, $n_q$ is a quantum density given by $(m\tau/2\pi\hslash^2)^{3/2}$ with dimensions of number density, and $V$ is the volume of the chamber.
In chemical equilibrium $\mu(h,\tau)=\mu(0,\tau)$, the number density is related to total number $N$ is given by
\begin{align}
n(h,\tau)&=n(0,\tau)\exp \left(\frac{-mgh}{\tau}\right),
\label{eqn:ce} \\ \ 
N&=s\int_0^L n(h,\tau) dh =\frac{n(0,\tau)\tau s}{mg}\left[1-e^{-mgL/\tau}\right].
\end{align}

From Gibbs-Duhem relation and chemical equilibrium eq.(\ref{eqn:ce}), the entropy density $\sigma(h,\tau)$ is given by 
\begin{align}
\sigma(h,\tau)&=-n(h,\tau) \left(\frac{d\mu}{d\tau}\right) +\left(\frac{dp}{d\tau}\right)\\
&=- n(h,\tau)[\log (n(0,\tau)/n_q)-5/2] \\
&=\sigma(0,\tau)(n(h,\tau)/n(0,\tau)),
\end{align}
substituting the pressure $p$ by $p=n\tau$.
Then total entropy is given by 
\begin{align}
S(\tau)=s\int_0^L \sigma(h,\tau) dh=\frac{\sigma(0,\tau) s}{n(0,\tau)}\int_0^L n(h,\tau)dh=\sigma(0,\tau) N/n(0,\tau).
\end{align}
At the merging of two black holes the uniform gravitational acceleration $g$ would be canceled out by the other uniform gravitational acceleration $-g$ of the opponent black hole.
Since we suppose that the merging entropy can be evaluated in the isothermal process, such entropy with gravitation would be evaluated at a merging temperature, Then we subtract it from an entropy $S'(\tau)$ which is evaluated after the merging where the system is without gravitation and in the same temperature (where the dash notation refers to values without gravitation). The uniform number density $n'=N/sL$ and entropy of the uniform ideal gas are related by $S'(\tau)=-N[\log (n'/n_q)-5/2]$.
Therefore the entropy increase by the offset of the uniform gravitation is given by
\begin{align}
\Delta S(\tau)=S'(\tau)-S(\tau)&=N \left[-\log (n'/n_q)+\log (n(0,\tau)/n_q)] \right]\\
&=N \log\frac{n(0,\tau)sL}{N} \\
&=N \log \frac{mgL}{\tau}\frac1{1-e^{-mgL/\tau}}, \\
\label{eqn:ds}
\end{align}
where $\tau$ is the merging temperature which will be estimated later.

Assuming that $m$, 
$mgL<<\tau$, which is justified with a large number of particles, eq.(\ref{eqn:ds}) accepts following approximation
\begin{align}
\Delta S(\tau)&\sim N \log \frac{mgL}{\tau} \frac1{\frac{-mgL}{\tau} (-1+ \frac{mgL}{2\tau}) }\\
&\sim N \log (1+\frac{mgL}{2\tau})\\
&\sim \frac{NmLg}{2 \tau}=\frac{NmLg}{2 k_B T}.
\end{align}
One can easily see that this relation is similar to the relation handled in the discussion of the entropic force\cite{V} except for the numerical coefficient.

Now we evaluate this formula for one of colliding black holes. 
The merging temperature would be estimated at Hawking temperature of each black hole$1/8\pi M_1$, while there might be some uncertainty.
The height of the cylinder chamber $L$ will roughly be a distance between the two black holes. In the present work, we evaluate this distance by the length of the crease set $l$\cite{CS}\cite{MS1}\cite{Ida}. 

The crease set is the subset of past endpoints of null horizon generators.    
As is pointed out in \cite{RP}, 
the {\em endpoint set}\ $\cal E$ of a horizon 
is an acausal subset (drawn in Fig.\ref{fig:sp}). Moreover in the spacetime where black holes settle
 into one spherical
black hole in the asymptotic future, $\cal E$ is arc-wise connected and homotope to a point.
Points $u\in\EPS$ are classified by 
the {\em multiplicity\/} $m(u)$ of $u$, 
the number of the generators emanating from $u$:
\begin{eqnarray}
  {\cal C}&:=&\{u\in\EPS\;|\;m(u)>1\},\nn
  {\cal D}&:=&\{u\in\EPS\;|\;m(u)=1\}.
\end{eqnarray}
The set $\cal C$ is called the {\em crease set} of the horizon. 
The crease set contains the interior of the endpoint set, i.e., the closure of $\cal C$ contains $\cal E$~\cite{CS}. 
The crease set $\cal C$ equals the set of points of $\cal E$ on which the 
horizon is not differentiable, i.e., 
the horizon is differentiable at $u\in\cal E$ if and only if
$u\in\cal D$~\cite{CS,Chr98CQG}.

 As studied in \cite{Ida}\cite{MS1}\cite{MS3}, the length of the crease set is useful to evaluate the initial separation between colliding black holes, since that is a covariant value of the distance determined by causal structure.
Especially the topology of event horizon strongly depends on the crease set and its timeslicing\cite{MS1}.
 Indeed, the black hole head-on collision is due to the line-like crease set as illustrated in the figure \ref{fig:sp}.
When we take a natural timeslicing for the black hole coalescence as in the figure \ref{fig:sp}, the spacetime points where two black holes are formed, are connected by the crease set. Therefore the length of the crease set $l$ would represent the distance between the two black holes as $L=l$.

Moreover, $g$ is roughly estimated by the surface gravity of the black hole $g=1/4M_1$. $Nm$ would be the mass of matter constituting the black hole.
Nevertheless, we are unsure the present evaluation is valid since the gravitational potential around each black hole would not be isotropic.

Rather we put a numerical factor $\alpha$ to parameterize that.
As dipole force is weaker than monopole force, $\alpha$ might be smaller than $1$. Moreover we will also include the uncertainty of the merging temperature in this numerical factor\footnote{$\alpha$ could include also other uncertainty since the argument to evaluate merging entropy was rather order estimation. Nevertheless we consider that these uncertainties are common to the low energy collision and the high energy collision in the leading order, and then we proceed the following discussion.}. In the following we calibrate the numerical factor $\alpha$ so as to explain the low energy collision where gravitational wave is hardly radiated. Then we will check its validity from the high energy collision.



Hence the black hole merging entropy is evaluated as
\begin{align}
\Delta S_1 =\frac{\alpha M_1 l}2  \frac1{4M_1}(8\pi M_1)=\alpha \pi M_1 l.
\label{eqn:me}
\end{align}
Total entropy increase is given by summing up the contributions of two black holes as $2\Delta S_1$.

\section{Energy ratio of gravitational radiation with merging entropy}
\begin{figure}[hbtp]
\includegraphics[width=7cm]{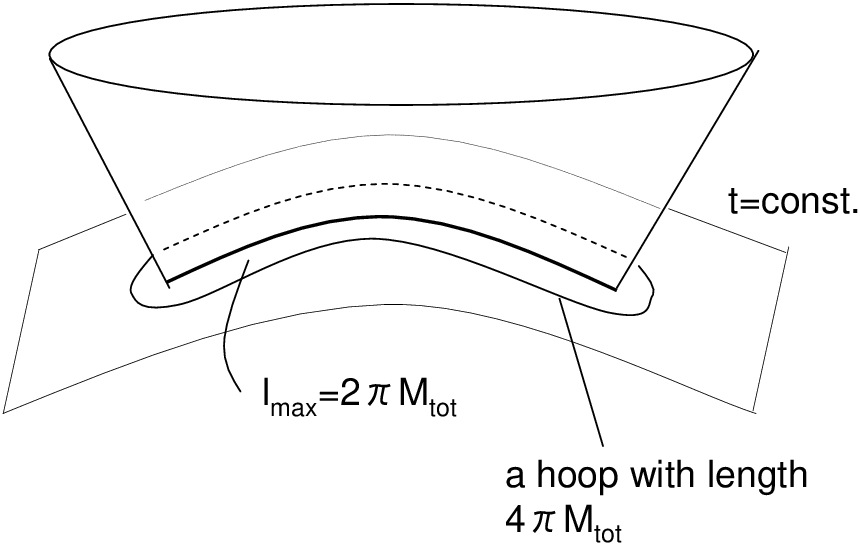}
\caption{The crease set implies a spindle-like black hole on an appropriate spatial hypersurface.
The crease set should be shorter than $2\pi M_{tot}$ so that a hoop with a length $4\pi M_{tot}$ can circulate the spindle-like black hole.}
\label{fig:hoop}
\end{figure}

Now we estimate the energy ratio of gravitational radiation by the area theorem of the event horizon incorporating the merging entropy.
Substituting the merging entropy (\ref{eqn:me}) into the area theorem 
 (\ref{eqn:AT3}) with the Bekenstein-Hawking relation $\Delta A_1=4\Delta S_1$, we have an inequality
\begin{align}
\frac{E}{2M_1}&<1-\sqrt{\frac12+\frac{\alpha l}{8 M_1}}.
\label{eqn:ie1}
\end{align}
Here we note that if we believe the hoop conjecture\cite{HO} the length of the crease set might be bounded from above. The hoop conjecture forbids such anisotropic event horizon that any hoop with length $4\pi M$ cannot surround it. On a certain spatial hypersurface (see Fig.\ref{fig:hoop}), a long crease set implies the existence of long spindle-like event horizon forbidden in the hoop conjecture\cite{Ida}.

In the case of low-energy collision, from the hoop conjecture we evaluate the length of the crease set $l$ by the maximum value $l_{max}=2\pi M_{tot}$ in order for a hoop with length $4\pi M_{tot}$ to circulate the spindle-like total black hole as shown in Fig.\ref{fig:hoop}.
For the total mass $M_{tot}=2M_1-E$, and the energy ratio $x=E/2M_1<1$, the inequality (\ref{eqn:ie1}) becomes
\begin{align}
x&<1-\sqrt{\frac12+\frac{2\alpha\pi(1-x)}{4}}.
\end{align}

In order for vanishing $x$, the numerical factor should be calibrated as $\alpha=1/\pi$.
In the following we use this numerical value.

Next we consider the case of high-energy collision of black holes. To estimate the energy ratio of the high-energy collision, we will discuss the length of crease set again.
Intuitively, we say that the black hole collision with short crease set corresponds to the high-energy collision since its spatial scale will be shortened by the Lorentz boost. Firstly we comment that the distinction between low- and high-energy collisions depends on timeslicing. By a coordinate transformation related to the Lorentz boost of the black holes, the irreducible mass $M_{irr}$ of the black hole can be converted into the kinetic energy brought by linear momentum $P$ while the ADM-energy of the colliding system does not change.
In the following we show that low-energy collision with identical masses $M_{irr}$ corresponds to high-energy collision with identical masses $M_{irr}'=\sqrt{M_{irr}^2-P^2}$ and linear momentum $P$, by an appropriate coordinate transformation which changes the coordinate separation between the colliding black holes.

\begin{figure}[hbtp]
\includegraphics[width=10cm]{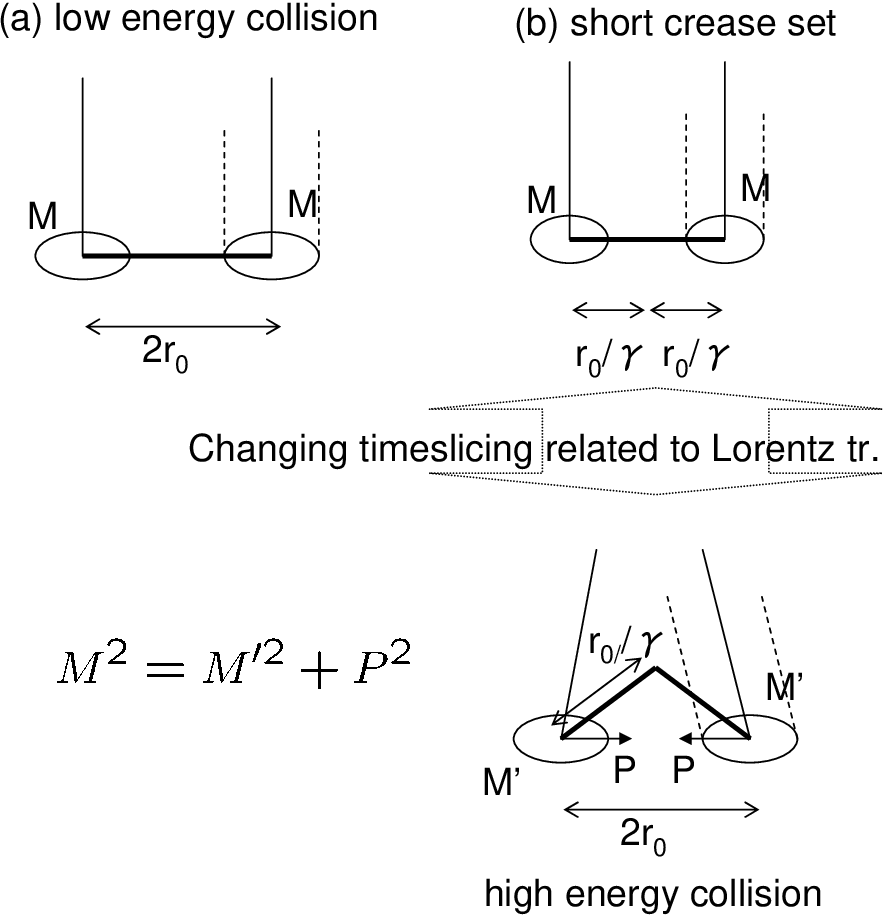}
\caption{On (a), the event horizon in the low-energy head-on collision is illustrated. The proper length of the crease set is $r_0$ and its total energy is $M$.  On (b), we prepare an event horizon which is similar to the event horizon in (a) but its crease set is with length $r_0/\gamma$. By coordinate transformation related to the Lorentz boost with factor $\gamma$, the coordinate separation becomes same as that of (a). That corresponds to the event horizon in the high-energy head-on collision as illustrated.}
\label{fig:bb}
\end{figure}

Here, to develop the same discussion between low- and high-energy collision, we prepare two pair of coalescing black holes whose crease sets are with length $2r_0\sim$Fig.\ref{fig:bb}(a) and with $2r_0/\gamma\sim$Fig.\ref{fig:bb}(b), where their crease sets are almost tangent to their timeslicing, respectively ($\gamma=1/\sqrt{1-\beta^2}, \beta=v/c$, and $v$ is velocity relative to mass center). To make coordinate separations between each colliding black holes same, we changes the timeslicing of collision (b) related to the Lorentz boost with factor $\gamma >1$ as illustrated in Fig.\ref{fig:bb}(b).

Then each black hole in (b) gains linear momentum (Bowen-York parameter\cite{BY}) $P=\sqrt{\gamma^2-1}M_{irr}$ and their irreducible mass $M_{irr}$ decreases by factor $1/\gamma$ since `ADM-energy' $M_{ADM}^2=M_{irr}^2+P^2$ does not change.
After all we see that when we compare two colliding systems with same ADM-energy and same coordinate separation the crease set of high-energy collision is shorter than that of low-energy collision in proper length.

Here it should be noted, however, even in very high-energy collision, the crease set cannot get as much shorter as possible.
Black holes too close to each other will not be regarded as the colliding of two black holes, because before their coalescence two black holes may already have been inside of a single black hole horizon.
Strictly speaking, this is correct for quasi-local horizons such as apparent horizons, not for event horizons. 
Since a spatial section of event horizon, however, must be surrounding the apparent horizon under certain physical conditions, two separated event horizons cannot exists there without another source of gravitation.
Then it seems that we cannot have such an initial conditions in vain.
We will estimate the minimal length of crease set for the very high-energy collision also by the hoop conjecture from this viewpoint.
 
\begin{figure}[hbtp]
\includegraphics[width=8cm]{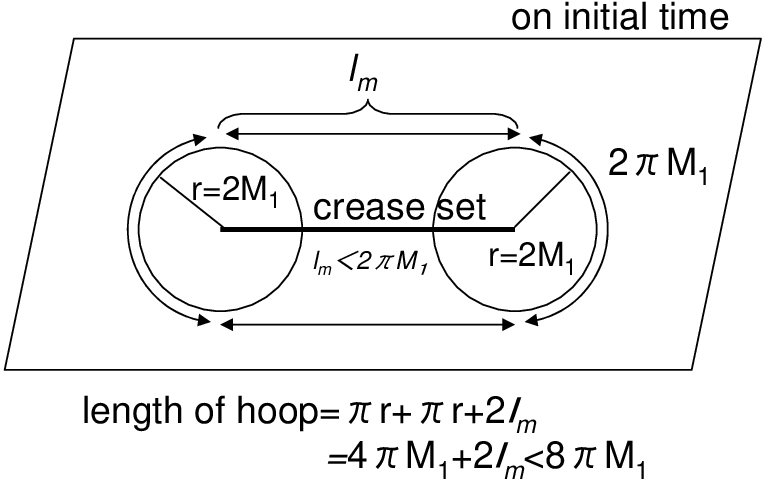}
\caption{Two half circles are connected as shown. Since this connected half circles with length $8\pi M_1$ circulates both colliding Schwarzschild horizons, initially there are not two colliding black holes but is only one merged black hole.}
\label{fig:hc}
\end{figure}

If two black holes are close to each other and their common crease set is shorter than $l_m=\pi (M_1+M_2)=2\pi M_1$, there would be a hoop surrounding the two black holes in a shape of connected two half circle (see Fig.\ref{fig:hc}) and with length $\pi (2M_1)+\pi (2M_1)+2l_m =8\pi M_1=4\pi (M_1+ M_2)$, which is the length of initial total mass $(M_1+M_2)$ multiplied by $4\pi$.
On a spatial hypersurface containing the crease set, the hoop conjecture implies that a common black hole horizon containing both two colliding black holes will have already existed.
Since this is not appropriate for initial condition for black holes to collide, we conclude $l>l_{min}=2\pi M_1$ for the high-energy black hole coalescence.

Hence, we evaluate the length of the crease set by $l=2\pi M_1$ for maximally boosted black hole collision.
The energy ratio $x=E/2M_1$ is given by the area theorem (\ref{eqn:ie1}) and $\alpha=1/\pi$ as
\begin{align}
\frac{E}{2M_1}&<1-\sqrt{\frac12+\frac{2 \alpha \pi M_1}{8 M_1}}\\
x&<1-\sqrt{\frac12+\frac14} \sim 0.134.
\end{align}

This upper bound is fairly close to an already given value 0.14 of high-energy limit in the numerical work\cite{HC} extrapolated by ZFL-PP calculation\cite{ZFL}\cite{PP} and the prediction of linear perturbation 0.16 by D'Eath and Payne in Ref.\cite{DP}. 
The fact that the area difference $\Delta A$ from the merging entropy $\Delta S$ explains simultaneously the numerical results of low-energy collision and the independent result of numerical simulation for high-energy collision by an appropriate choice of the numerical factor $\alpha$ implies one would be able to complete consistent discussion using the merging entropy.
 From that, we might conclude that the dynamical effect of collision is always represented by this entropy increase of merging process.

Furthermore we investigate $\gamma$-factor dependence in small $\beta$ limit by the Lorentz contraction of the crease set.
From inequality (\ref{eqn:ie1}) and $l=l_{max}/\gamma=2\pi M_{tot}/\gamma$, the energy ratio is bounded by
\begin{align}
\frac{E}{2M_1}&<1-\sqrt{\frac12+\frac{M_{tot}}{4 M_1}\frac1{\gamma}}\\
\Rightarrow \ \ \  x&<1-\frac1{4\gamma}-\sqrt{\frac12+\frac1{16\gamma^2}}\sim \frac16\beta^2+\frac7{216}\beta^4... .
\end{align}
There is no correspondence to the formula of ZFL-PP approximation in Ref.\cite{ZFL}\cite{PP}. Especially the $\beta^2$ dependence is also not reported by the numerical result.
Probably, that will be caused by $\gamma$-dependence in the large $\gamma$ limit, which is hard to estimate since the length of the crease set should be saturated at $l=2\pi M_1$ by above intuitive discussion.

\section{conclusion and discussions}
We have studied the amount of energy converted into gravitational radiation in black hole head-on collision, revising Hawking's discussion by incorporating the entropy increase by merging.
In the head-on collision of identical mass black holes, introducing the merging entropy which resembles mixture entropy, area theorem of the event horizon is re-evaluated. Incorporating the re-evaluated area theorem to Hawking's original discussion to have an upper bound of energy ratio of the gravitational radiation, the upper bound is lowered from Hawking's 29\% to negligible ($\sim$ 0\%) in low-energy collision and to 13.4\% in maximally high-energy limit simultaneously by tuning the numerical factor which adjusting anisotropic aspects.
These results well agree with two independent numerical simulations which are 0.1\% in low-energy\cite{NA} and 14\% in maximally high-energy limit\cite{HC}, at once.
Nevertheless there was no clear ground for identifying the merging entropy and the entropy of the toy-model of gas. Furthermore, as we have avoided precise estimation, e.g., our discussion includes unreliable evaluation for a rigorous numerical coefficient, the upper bounds would be typical values rather than strict upper bounds.

Therefore our discussion is not good for astrophysical prediction, while one may be hopeful that the merging entropy is useful for study coalescing process of the black hole.
Of course, in real collision of the black holes total entropy would be much increasing by thermal process of matter field, which is significant for the second low of thermodynamics. Nevertheless, the present discussion is based not on the increasing low of the total entropy but on the increasing low of the black hole event horizon area. The corresponding entropy is considered to be only black hole entropy related to the horizon area. 
From the present study of the black hole collision, we would convince there is a remarkable contribution of the merging entropy to the black hole entropy (or the area of the event horizon), which can be related to the topology of the event horizon.
On the other hand, since the numerical factor is chosen to be $1/\pi$ composed of fundamental geometrical number, it might be possible to derive that numerical factor by a purely geometrical calculation.

It is important question whether the method also correctly predict the energy
flux  from unequal mass head-on collision or spinning black hole collision.
For unequal mass we will find extra parameters of anisotropy to be fixed,
since gravitational force contains higher order anisotropic components.
Also for spinning black holes, we will find additional parameters, related
to the deformation of event horizon by their spin.
To fix such additional parameters, we need many data of numerical collision.
If the numerical study remarkably progresses, there would be a chance to develop
the method used in the present study.

In the present analysis, the contribution of the merging entropy implies the merging process is not an invertible process. That is consistent with the fact that black holes can merge but never split.
In the black hole dynamics, is there another not invertible process? If so, we will find another component of the black hole entropy like the merging entropy.
By contraries, one may speculate that more gravitational radiation emitted during the transition from toroidal black hole to a spherical black hole since such process is topologically invertible\cite{MS1}\cite{IdaS}. In such an invertible process, it can not be expected that any dynamical component of the black hole entropy lower the upper bound of the energy ratio.
Moreover that would be consistent with the fact that in circular orbiting coalescence or black hole scattering with impact parameter\cite{BSC} of the black holes more gravitational radiation can be emitted, because they would be regarded as the formation and collapse of the toroidal black hole\cite{SIda}.

In microscopic scales described by quantum gravity, it might be possible for a black hole to split. Are there any failures of our analysis in that sub-Planckian geometry?
This is probably understood as to be the following. The area theorem of black hole requires an energy condition. Since in such microscopic situations quantum effect will violate the energy condition, our analysis of the merging entropy loses its foundation and is consistent with the fact that this process becomes invertible.

Finally, can these tools be useful also in higher-dimensional scenarios? In five-dimensional collision\cite{5D}, the upper bound predicted from Hawking's area theorem is lowered and the numerical result is larger than that of four-dimensional collision.
Since area ratio of sphere and cone in five-dimensions is smaller than in four-dimensions, it seems possible to explain the dimensional dependence by the area differences. With numerical results of the high energy collision, we would rearrange the factor of ambiguity $\alpha$ and  discuss the validity of the merging entropy.

\end{document}